# First-principles Nonadiabatic Dynamics of Molecules at Metal Surfaces with Vibrationally Coupled Electron Transfer

Gang Meng[1], James Gardner[2], Wenjie Dou[3], Reinhard J. Maurer[2,4,*], Bin Jiang[1,5*]

[1]Key Laboratory of Precision and Intelligent Chemistry, Department of Chemical Physics, University of Science and Technology of China, Hefei, Anhui, China.
[2]Department of Chemistry, University of Warwick, Coventry CV4 7AL, United Kingdom.
[3]Department of Chemistry, School of Science, Westlake University, Hangzhou 310024 Zhejiang, China.
[4]Department of Physics, University of Warwick, Coventry CV4 7AL, United Kingdom.
[5]Hefei National Laboratory, University of Science and Technology of China, Hefei, 230088, China.

Accurate description of nonadiabatic dynamics of molecules at metal surfaces involving electron transfer has been a longstanding challenge for theory. Here, we tackle this problem by first constructing high-dimensional neural network diabatic potentials including state crossings determined by constrained density functional theory, then applying mixed quantum-classical surface hopping simulations to evolve coupled electron-nuclear motion. Our approach accurately describes the nonadiabatic effects in CO scattering from Au(111) without empirical parameters and yields results agreeing well with experiments under various conditions for this benchmark system. We find that both adiabatic and nonadiabatic energy loss channels have important contributions to the vibrational relaxation of highly vibrationally excited CO($v_i$ = 17), whereas relaxation of low vibrationally excited states of CO($v_i$ = 2) is weak and dominated by nonadiabatic energy loss. The presented approach paves the way for accurate first-principles simulations of electron transfer mediated nonadiabatic dynamics at metal surfaces.

Energy transfer during gas-surface collisions plays a crucial role in many interfacial phenomena [1]. Various degrees of freedom (DOFs) - molecular vibration, rotation, and translation, along with surface phonons and electrons - can potentially couple to each other, leading to intricate dynamics of energy transfer. Especially at metal surfaces, the energy threshold for electronic excitations in metals is infinitesimally small rendering nonadiabatic energy exchange between molecular motion and metallic electrons inevitable. Experiments of adsorbate vibrational lifetimes [2], chemicurrents [3], and hydrogen scattering [4] have provided clear evidence of nonadiabatic effects at metal surfaces. These effects have been largely understood by electronic friction (EF) based models [5,6], in which weak nonadiabatic energy dissipation is treated perturbatively [4,7-11].

In contrast, the scattering of vibrationally excited NO and CO molecules from metal surfaces manifests diverse vibrational energy transfer effects that depend on molecule and substrate [12-15], the initial vibrational state [16,17], and the molecular orientation [16,18]. The observed multi-quantum vibrational relaxation is associated with vibrationally enhanced electron transfer from metals to molecules [12] — a strong nonadiabatic effect involving the formation of a transient ionic state as qualitatively interpreted by low-dimensional phenomenological models [19]. Meanwhile, recent Born-Oppenheimer molecular dynamics (BOMD) studies based on first-principles machine-learned potential energy surfaces (PESs) indicate that vibrational energy may transfer efficiently to other low frequency modes, as the molecule approaches the dissociation barrier region [20-22]. On top of an adiabatic PES, MD simulations with EF based on density functional perturbation theory can capture the single-quantum and orientation-dependent vibrational relaxation in the scattering of NO($v$ = 3) from Au(111) [23,24], but largely underestimate the multi-quantum vibrational relaxation and overestimate molecular trapping for highly vibrationally excited NO [23].

To better describe these events, it is essential to accurately capture both the high-dimensional PES and nonadiabatic electronic transitions. Among methods for nonadiabatic dynamics near metal surfaces that go beyond the classical EF theory [25-28], the independent electron surface hopping (IESH) [25] approach is the most used one [29-31] and so far the only one that has been invoked to describe high dimensional dynamics of a realistic system based on electronic structure theory, namely for NO on Au(111), with some successes [25,32-34]. This method takes an independent electron representation and allows for single electron hopping within a manifold of electronic states that couple discretized metallic and molecular levels in a mixed quantum-classical manner. Unfortunately, it later turned out that the parameterization of diabatic states constructed with an effective perturbative approach based on an applied electric field failed to capture important features of the dynamics. This resulted in disagreements with experiments under various conditions [17,24,35].

We recently proposed a more general way to calculate charge-transfer states of molecules at metal surfaces by constrained density functional theory (CDFT), which successfully described the electron transfer behavior of NO and CO on different metal surfaces [36]. In this Letter, we use a high-dimensional machine-learning algorithm to represent CDFT energies at varied nuclear configurations for the construction of an effective diabatic Hamiltonian for IESH, enabling a first-principles description of nonadiabatic dynamics at metal surfaces mediated by electron transfer. As a proof of concept, we study the CO + Au(111) system, in which the nonadiabatic effects are subtle and both adiabatic and nonadiabatic energy transfer channels need to be accurately captured [14,37].

The IESH method originates from a discretized version of the Newns–Anderson Hamiltonian [38] considering only two molecular electronic configurations, *i.e.*, the neutral and negative ion states. In this model, the many-electron Hamiltonian is expressed as a sum of one-electron terms given by [25,31],

$$H(\mathbf{R},\mathbf{P}) = \sum_{i=1}^{3N} \frac{P_i^2}{2M_i} + U_0(\mathbf{R}) + \sum_{j \in \mathbf{s}(t)} E_j(\mathbf{R}), \quad (1)$$

where the first term is the nuclear kinetic energy and the third term is the sum of energies of occupied one-electron orbitals indexed in the time-dependent $\mathbf{s}$ vector. $E_j(\mathbf{R})$ is the $j$th orbital energy or the $j$th eigenvalue of the one-electron Hamiltonian,

$$H_{el}^1(\mathbf{R}) = (U_1(\mathbf{R}) - U_0(\mathbf{R}))|a\rangle\langle a| + \sum_{k=1}^{M} \varepsilon_k |k\rangle\langle k| \\ + \sum_{k=1}^{M} V_{ak}(\mathbf{R})(|a\rangle\langle k| + |k\rangle\langle a|) \quad , \quad (2)$$

where $U_0(\mathbf{R})$ (or $U_1(\mathbf{R})$) represents the interaction potential between the metal surface and the neutral molecule (or the negative ion when the lowest unoccupied orbital of the molecule, $|a\rangle$, is occupied). The metallic continuum is discretized by $M$ one-electron orbitals $\{|k\rangle\}$ corresponding to one-electron energies of $\{\varepsilon_k\}$. $V_{ak}(\mathbf{R})$ represents the coupling strength between $|a\rangle$ and $|k\rangle$, which is often assumed to be constant over the relevant energy range. In this representation, the ground state corresponds to filling the $N_e$ ($N_e = M/2$) lowest energy one-electron eigenstates up to the Fermi level, while excited states are produced by promoting one or more electrons to unoccupied states. One then employs the modified "fewest switches" surface hopping algorithm [39] to model transitions of multiple independent electrons [25]. More details of the IESH method are given in the Supplemental Material (SM, Figs. S1-S3).

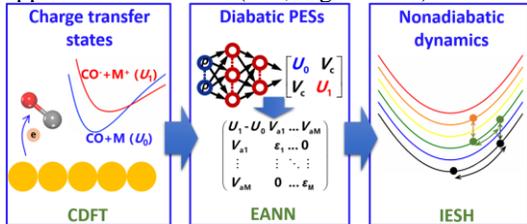

FIG. 1. Schematic workflow of nonadiabatic dynamics simulations of molecules (CO for example) at metal surfaces involving electron transfer between them.

Here, we propose to parameterize the IESH Hamiltonian by the CDFT-based diabatic states and corresponding machine learning potentials as illustrated in Fig. 1. There are several critical improvements of this workflow over the initial IESH application to the NO + Au(111) system [25,32,33]. First, $U_0(\mathbf{R})$ (or $U_1(\mathbf{R})$) is determined by CDFT, for which a net Bader charge of 0 $e$ or -1 $e$ is constrained to the molecule by adjusting self-consistently an external potential in the modified Kohn-Sham equations [40,41]. As depicted in Fig. S4, CDFT is more stable than the previous electric field-based perturbative method [32], ensuring better asymptotic behavior and global smoothness of diabatic states. Second, to overcome the failure of previously used empirical functions to describe molecular dissociation and anharmonicity of the lattice vibration [32], we utilize a high-fidelity embedded atom neural network (EANN) method [42,43] to learn CDFT energies, yielding high-dimensional diabatic PESs that are as accurate as the adiabatically ground state ($E_g$) PES learned from conventional DFT energies and forces. The off-diagonal coupling ($V_c$) between the two diabatic states can be derived by enforcing consistency between $E_g$ and the lowest eigenvalue of the 2×2 diabatic Hamiltonian. $V_c$ is then used to parameterize $V_{ak}(\mathbf{R})$ in Eq. (2) (see SM [25]. Specifically, CDFT calculations for CO + Au(111) were performed with CP2K [44] in a slab model of a 6×6 supercell with four metal layers using the van der Waals density functional (vdW-DF) [45] — the same level of theory used for $E_g$ [22]. As shown in Fig. 2, although $U_1$ is generally higher in energy than $U_0$ at most area, the two diabats do cross at very short molecular height and very long C-O distance, which plays an important role in nonadiabatic energy transfer. EANN potentials nicely reproduce CDFT (and ground state DFT) energies, capturing the crossing between diabatic states. More details and the validation of ground and diabatic PESs are given in SM (Figs. S5-S6).

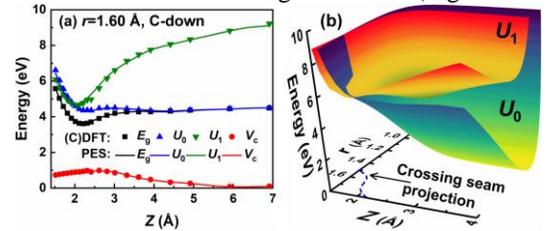

FIG. 2. (a) Potential energy curves of the ground state ($E_g$), the neutral ($U_0$), anionic state ($U_1$), and the diabatic coupling ($V_c$) as a function of the molecular height ($Z$) above Au(111). CO is perpendicular to the hcp site with a long bond distance ($r$ = 1.60 Å) and C-down orientation. (b) Diabatic state PESs as a function of $r$ and $Z$, where the crossing seam of the diabatic states is projected onto the $r$-$Z$ plane (dashed line). The energy zero is $E_g$ of the free CO molecule far from the surface.

We apply this approach to study state-to-state scattering of CO from Au(111), where extensive experimental data exists [14,46,47]. First, Fig. 3 compares the final vibrational state distributions of highly vibrationally excited CO($v_i$ = 17) scattered from Au(111) calculated by BOMD and IESH simulations, with experimental data at three translational incidence energies ($E_i$) [14]. The experimental survival probability of CO($v_i$ = 17) decreases as $E_i$ increases from 0.24 eV to 0.57 eV and the vibrational state relaxes down to $v_f$ = 14. BOMD predicts the proper $E_i$-dependence of vibrational relaxation, yet overestimates the survival probability by 0.1 – 0.2. The promoted vibrational energy loss for molecules at higher $E_i$ is attributable to their higher accessibility to the dissociation barrier region where the molecular vibration softens and couples to translation [21,22]. In comparison, IESH predicts more significant vibrational relaxation and brings the calculated vibrational state distributions close to measured ones. Additionally, IESH results capture the multi-quantum vibrational relaxation feature, although the probability is relatively low, while BOMD results are dominated by single-quantum relaxation. A representative IESH trajectory in Fig. S7 shows that more than 0.5 eV of vibrational energy (~ two quanta) can rapidly dissipate to surface electrons within ~80 fs via a single bounce between the molecule and the surface.

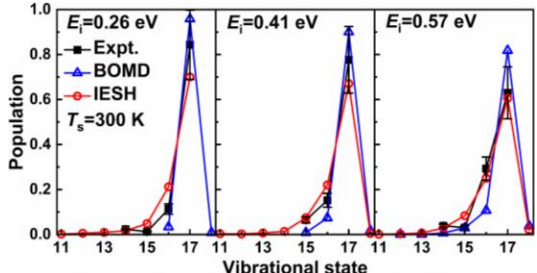

FIG. 3. Comparison of experimental final vibrational state distributions [14] of CO($v_i$ = 17) scattered from Au(111) at different incidence energies with BOMD and IESH results based on the vdW-DF PES. The surface temperature ($T_s$) is 300 K, as in experiments.

To elaborate different energy transfer pathways, Table I summarizes the mean energy loss (or gain) of molecular vibration (<$\Delta E_{vib}$>), rotation (<$\Delta E_{rot}$>), translation (<$\Delta E_{trans}$>), surface phonons (<$\Delta E_{ph}$>), and surface electrons (<$\Delta E_{el}$>) during CO($v_i$ = 17) scattering at $E_i$ = 0.57 eV. BOMD results show that the translational energy of CO is largely transferred to rotation (~0.12 eV) and phonons (~0.24 eV) during vibrationally elastic scattering. In comparison, for vibrationally inelastic scattering, <$\Delta E_{trans}$> is greatly increased by ~0.18 eV that well matches the vibrational energy loss, implying that vibration is mainly coupled to translation in this process. Experiments have also indicated a certain degree of vibration-to-translation coupling in NO scattering from Au(111) [48]. Interestingly, IESH simulations barely change the energy exchange among other DOFs, except that <$\Delta E_{trans}$> is now ~0.11 eV lower than that in BOMD for vibrationally inelastic scattering. On average, vibrational energy is primarily transferred to surface electrons (~0.19 eV), secondarily to translation (~0.09 eV), while negligibly to rotation and surface phonons. Comparison of representative trajectories of BOMD and

IESH with identical initial conditions in Fig. S7 clearly shows that the energy variation in translation, rotation, and surface phonons is hardly affected by considering energy dissipation due to electronic excitation. In turn, the lattice motion has little influence on vibrational energy transfer in BOMD and IESH simulations (Fig. S8). These results indicate that, under these conditions and for this system, the adiabatic and nonadiabatic pathways of vibrational energy transfer are rather independent and both represent additive contributions. The latter always dominates but the former cannot be neglected.

TABLE I. Average vibrational, rotational, translational, surface phonon and surface electron energy losses (or gains) (in eV) of CO($v_i$ = 17, $E_i$ = 0.57 eV) scattering from Au(111) in BOMD and IESH simulations. Vibrationally elastic ($v_f$ = 17) and inelastic ($v_f \neq$ 17) channels are separately listed.

| Mean energy change (eV) | BOMD | | IESH | |
|---|---|---|---|---|
| | $v_f$ = 17 | $v_f \neq$ 17 | $v_f$ = 17 | $v_f \neq$ 17 |
| <$\Delta E_{vib}$> | -0.01 | -0.17 | -0.01 | -0.28 |
| <$\Delta E_{rot}$> | 0.12 | 0.07 | 0.13 | 0.09 |
| <$\Delta E_{trans}$> | -0.35 | -0.17 | -0.36 | -0.28 |
| <$\Delta E_{ph}$> | 0.24 | 0.27 | 0.24 | 0.28 |
| <$\Delta E_{el}$> | 0.00 | 0.00 | 0.00 | 0.19 |

Next, we focus on the scattering of CO in a low vibrational state ($v_i$ = 2) from Au(111), which represents a benchmark process that exhibits weak nonadiabaticity [46,47]. As shown in Fig. 4(a), the vibrational inelasticity of CO($v_i$ = 2) is much weaker than that of CO($v_i$ = 17). Moreover, the vibrational relaxation probability first decreases and then increases with increasing $E_i$. Notably, BOMD results fail to predict any vibrational inelasticity. In contrast, present IESH results reproduce the small but finite relaxation probability for CO($v_i$ = 2 → $v_f$ = 1) within a factor of two compared to experiments and its $E_i$-dependence. This signifies that the vibrational motion of CO($v_i$ = 2) couples solely, albeit weakly, to surface electrons, which is essential to trigger the vibrational relaxation. Indeed, molecules in this lower state cannot access the dissociation region so that the adiabatic vibration-to-translation coupling becomes negligible. Importantly, we find that vibrational relaxation arises mostly from direct scattering (DS) trajectories for 0.48 eV ≤ $E_i$ ≤ 0.92 eV. In this range, the increasing incidence energy renders the molecule closer to the strong coupling region near the surface resulting in more facile electron hopping. While at $E_i$ = 0.25 eV, which is close to the physisorption well depth of CO on Au(111) (0.19 eV), around 60% of inelastic CO($v_f$ = 1) products are resulted from those trajectories being temporarily trapped before desorption (TD) and experiencing an increased chance of electron hopping. Fig. S9 demonstrates the representative DS and TD trajectories with fast and slow electron hopping, respectively. Additionally, it is found that the positions for occurring electron hopping are generally farther from the surface at $E_i$ = 0.25 eV than at $E_i$ = 0.92 eV, and dominated by the TD component near the physisorption well. This higher TD contribution facilitates the vibrational coupling to surface electrons, thereby explaining the increased relaxation probability at $E_i = 0.25$ eV.

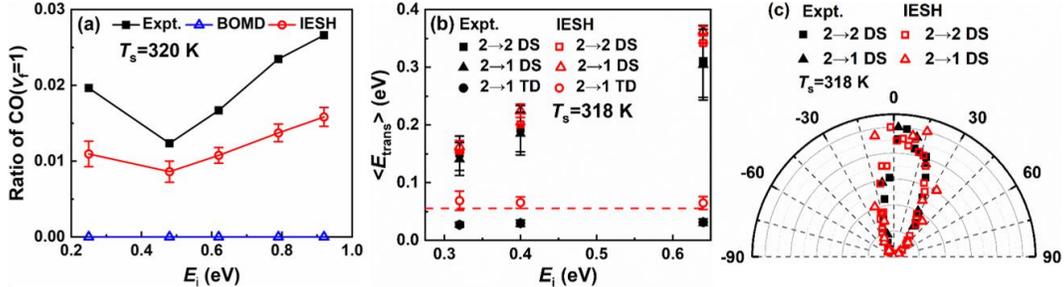

FIG. 4. Comparison of various experimental [46,47] and calculated results on the vdW-DF PES for the scattering of CO($v_i = 2$) from Au(111), including (a) product branching ratio of CO($v_f = 1$) and (b) final mean translational energy ($<\Delta E_{trans}>$) of different channels as a function of $E_i$, and (c) angular distributions for vibrationally elastic (2→2) and inelastic (2→1) channels of the DS component at $E_i = 0.32$ eV. Red dashed line in panel (b) indicates the thermal limit, i.e., $<\Delta E_{trans}> = 2k_bT_s$. Error bars reflect standard errors due to finite trajectories at each condition.

The IESH results also agree with experimental data of mean translational energy ($<\Delta E_{trans}>$) and angular distributions of scattered CO molecules [46], as shown in Figs. 4(b)-(c). Specifically, CO($v_f = 2$) and CO($v_f = 1$) molecules obtained from the DS channel have very close $<\Delta E_{trans}>$, both of which increase linearly with $E_i$. This indicates some memory of their initial conditions and confirms that vibrational energy loss to translation is negligible. They also share similarly narrow scattering angular distributions peaking near the specular angle, as expected for a DS process. By contrast, the $<\Delta E_{trans}>$ of TD component of CO($v_f = 1$) is much smaller and close to the thermal limit, i.e., $2k_bT_s$, where $T_s$ is the surface temperature of 318 K. It is nearly independent of $E_i$, suggesting that these molecules are largely equilibrated with the surface before desorption. Quantitatively, the calculated $<\Delta E_{trans}>$ of CO($v_f = 1$) molecules in the TD channel is slightly higher than the thermal limit as trapped trajectories are identified here by the maximum simulation time (50 ps), where thermalization may be incomplete.

Finally, Fig. S10 shows that previous BOMD results [22] using an alternate PES based on the BEEF-vdW functional [37], which gives a dissociation barrier that is ~0.5 eV lower than the vdW-DF PES, heavily overestimate vibrational relaxation for CO($v_i = 17$) compared to experiments. One would expect even larger discrepancy with experiments using this PES if nonadiabatic effects were accounted for. By contrast, this BEEF-vdW PES predicts negligible vibrational relaxation probabilities (~$10^{-4}$) for CO($v_i = 2$) when simulated with BOMD and when nonadiabatic effects are considered by an EF model within the local density approximation [37], in qualitative disagreement with experiments (see Fig. S11 and Ref. [37]). This indicates that vibrational energy transfer is quite sensitive to the precise energy landscape in the barrier region, which is a crucial prerequisite for the correct non-perturbative treatment of nonadiabatic gas-surface dynamics.

To summarize, we propose a novel and general computational scheme to enable a first-principles description of nonadiabatic dynamics of molecules at metal surfaces involving explicit electron transfer. In this scheme, charge-transfer states are determined by CDFT and full-dimensional diabatic PESs are represented by EANN, which are then integrated with the IESH algorithm allowing for energy flow to both surface phonons and electrons. Taking the CO + Au(111) system as an example, the simulation results achieve best agreement so far with experiments for both CO($v_i = 17$) and CO($v_i = 2$) scattering events from Au(111). Moreover, it is found that in the case of CO($v_i = 17$), vibration is predominantly coupled nonadiabatically to surface electrons, followed by adiabatic coupling to translation. While in the case of CO($v_i = 2$), vibration becomes almost exclusively coupled to surface electrons so that BOMD is unable to predict any vibrational inelasticity. Our results emphasize that an accurate description of vibrational energy transfer requires that both the adiabatic PES and nonadiabatic effects are accurately represented. This work introduces new opportunities for studying high-dimensional nonadiabatic dynamics at metal surfaces. In this respect, the CDFT-based diabatic state PESs can be integrated with other mixed quantum-classical algorithms such as the (broadened) classical master equation (BCME) [27]. An effective EF tensor can be also reformulated by the same many-electron Hamiltonian [34] for more efficient MDEF simulations. It will be interesting to compare the performance of these methods on the same first-principles basis in the future.

This work is supported by the Strategic Priority Research Program of the Chinese Academy of Sciences (XDB0450101), Innovation Program for Quantum Science and Technology (2021ZD0303301), National Natural Science Foundation of China (22325304 and 22221003), CAS Project for Young Scientists in Basic Research (YSBR-005), the Leverhulme Trust (RPG-2019-078), the UKRI Future Leaders Fellowship program (MR/X023109/1), and a UKRI Frontier research grant (EP/X014088/1). Calculations have been done on the supercomputing center of USTC and Hefei Advanced Computing Center. We thank Alec Wodtke,